\documentclass[acmsmall]{acmart}
\AtBeginDocument{%
  \providecommand\BibTeX{{%
    \normalfont B\kern-0.5em{\scshape i\kern-0.25em b}\kern-0.8em\TeX}}}

\usepackage{subfigure}
\usepackage{graphicx}

\usepackage{enumitem}
\usepackage{multirow}
\usepackage{makecell}
\usepackage{amsmath}
\usepackage{amsfonts}
\usepackage{url}

\newcommand{\ie}{\emph{i.e., }}
\newcommand{\eg}{\emph{e.g., }}
\newcommand{\etal}{\emph{et al. }}
\newcommand{\etc}{\emph{etc.}}
\newcommand{\wrt}{\emph{w.r.t. }}

\newcommand{\aka}{\emph{aka. }}

\newcommand{\yz}[1]{{\color{black}{#1}}}

\setcopyright{acmcopyright}
\copyrightyear{2024}
\acmYear{2024}
\acmDOI{XXXXXXX.XXXXXXX}

\acmJournal{TOIS}

\begin{document}

\title{MultiCBR: Multi-view Contrastive Learning for Bundle Recommendation}

\author{Yunshan Ma}
\email{yunshan.ma@u.nus.edu}
\authornote{Equal contribution.}
\affiliation{%
  \institution{National University of Singapore}
  \city{Singapore}
  \country{Singapore}}

\author{Yingzhi He}
\email{heyingzhi@u.nus.edu}
\authornotemark[1]
\affiliation{%
  \institution{National University of Singapore}
  \city{Singapore}
  \country{Singapore}}

\author{Xiang Wang}
\email{xiangwang1223@gmail.com}
\authornote{Corresponding author. Xiang Wang is also affiliated with Institute of Artificial Intelligence, Institute of Dataspace, Hefei Comprehensive National Science Center. This research is supported by the National Natural Science Foundation of China (9227010114) and the University Synergy Innovation Program of Anhui Province (GXXT-2022-040). This research is also supported by the National Natural Science Foundation of China (62172226), and the 2021 Jiangsu Shuangchuang (Mass Innovation and Entrepreneurship) Talent Program (JSSCBS20210200). This research is also supported by NExT Research Center.}
\affiliation{%
  \institution{University of Science and Technology of China}
  \city{Hefei}
  \country{China}}

\author{Yinwei Wei}
\email{weiyinwei@hotmail.com}
\affiliation{%
  \institution{Monash University}
  \city{Melbourne}
  \country{Australia}}

\author{Xiaoyu Du}
\email{duxy@njust.edu.cn}
\affiliation{%
  \institution{Nanjing University of Science and Technology}
  \city{Nanjing}
  \country{China}}

\author{Yuyangzi Fu}
\email{yfuyu@ebay.com}
\affiliation{%
  \institution{eBay Inc.}
  \city{Shanghai}
  \country{China}}
  
\author{Tat-Seng Chua}
\email{dcscts@nus.edu.sg}
\affiliation{%
  \institution{National University of Singapore}
  \city{Singapore}
  \country{Singapore}}


\begin{abstract}
Bundle recommendation seeks to recommend a bundle of related items to users to improve both user experience and the profits of platform. Existing bundle recommendation models have progressed from capturing only user-bundle interactions to the modeling of multiple relations among users, bundles and items. CrossCBR, in particular, incorporates cross-view contrastive learning into a two-view preference learning framework, significantly improving SOTA performance. It does, however, have two limitations: 1) the two-view formulation does not fully exploit all the heterogeneous relations among users, bundles and items; and 2) the "early contrast and late fusion" framework is less effective in capturing user preference and difficult to generalize to multiple views. 

In this paper, we present MultiCBR, a novel \textbf{Multi}-view \textbf{C}ontrastive learning framework for \textbf{B}undle \textbf{R}ecommendation. First, we devise a multi-view representation learning framework capable of capturing all the user-bundle, user-item and bundle-item relations, especially better utilizing the bundle-item affiliations to enhance sparse bundles' representations. Second, we innovatively adopt an "early fusion and late contrast" design that first fuses the multi-view representations before performing self-supervised contrastive learning. In comparison to existing approaches, our framework reverses the order of fusion and contrast, introducing the following advantages: 1) our framework is capable of modeling both cross-view and ego-view preferences, allowing us to achieve enhanced user preference modeling; and 2) instead of requiring quadratic number of cross-view contrastive losses, we only require two self-supervised contrastive losses, resulting in minimal extra costs. Experimental results on three public datasets indicate that our method outperforms SOTA methods. The code and dataset can be found in the github repo \url{https://github.com/HappyPointer/MultiCBR}.
\end{abstract}

\ccsdesc[500]{Information systems~Recommender systems}


\keywords{Bundle Recommendation, Graph Neural Network, Contrastive Learning}

\maketitle

\section{Introduction} \label{sec:introduction}
As a popular marketing strategy, item bundling has been gradually adopted by various online services, such as composing a set of compatible fashion items into an outfit or creating a music list by including songs of similar styles. Each individual item gains more opportunities to be exposed in diverse bundles as a result of item bundling. Consequently, indecisive consumers may be persuaded to purchase a specific item due to the justification provided by the innovative bundles. Because of these benefits, bundle recommendation has been proposed recently and received a lot of attention.

Early studies on bundle recommendation~\cite{FPMC2010} simplify it as a special type of user-item recommendations and directly adopt Collaborative Filtering (CF)-based methods, to capture the user-bundle interaction patterns. However, such simplification ignores the fact that a bundle is not an atomic unit but rather a collection of fine-grained elements, \ie individual items. Following works~\cite{cao2017embedding,DAM2019,BundleNet2020,BGCN2020,MIDGN2022,CrossCBR2022} take into account the user-item interaction and bundle-item affiliation information in order to take advantage of the valuable information carried by the items. Factorization models~\cite{cao2017embedding,DAM2019}, multi-task learning~\cite{cao2017embedding,DAM2019,BundleNet2020}, and Graph Neural Networks (GNNs)~\cite{BundleNet2020,BGCN2020,MIDGN2022,CrossCBR2022} are typical techniques and have been shown to be effective in dealing with multiple relations among users, bundles, and items. BGCN~\cite{BGCN2020}, in particular, sorts user preference over bundles into two views: bundle view (represents preference for the entire bundle) and item view (represents preference for the items within bundles), which is insightful in guiding advanced model design. By explicitly modeling the cooperative association between the two different views with cross-view contrastive learning, CrossCBR~\cite{CrossCBR2022} further improves the performance. Despite the impressive performance improvement, there are still a few limitations.

\begin{figure}[t!]
    \centering
    \includegraphics[width = 0.65\linewidth]{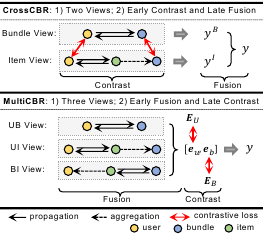}
    \caption{Brief comparison between MultiCBR and CrossCBR: 1) MultiCBR extends two views to three views, where the additional BI view aims to capture the bundle composition patterns implied by BI graph; and 2) MultiCBR adopts an "early fusion and late contrast" framework, which can better model user preference while introducing few extra expenses.}
    \label{fig:motivation}
    \vspace{-15pt}
\end{figure}

First, the formulation of two-view preference does not fully exploit the multiple heterogeneous relations among users, items, and bundles. As shown in Figure~\ref{fig:motivation}, CrossCBR performs \textit{bi-directional} information propagation over the user-bundle (UB) and user-item (UI) graphs, while applying \textit{uni-directional} aggregation on the bundle-item (BI) graph to garner information from item to its affiliated bundle.
However, we argue that this uni-directional aggregation is insufficient to make full use of the BI graph, such as the crucial patterns for bundle composition.
More importantly, for items and bundles that have fewer connections, the collaborative effects in the BI graph can significantly enhance these sparse nodes' representations. Consequently, fully exploiting the BI graph is an indispensable part of bundle recommendation (See the evidence in Section \ref{subsec:model_study}).

Second, but perhaps more important limitation is an "early contrast and late fusion" design in CrossCBR, neglecting the cross-view user preference and restricting it from generalizing to more complicated scenarios such as multiple views (\eg auxiliary information such as content or other relational data). As shown in Figure~\ref{fig:motivation}, CrossCBR just models user preference within each view, \ie ego-view preference of $y^B$ and $y^I$, while overlooking the potential cross-view preference. In addition, it requires constructing contrastive losses between every pair of views. Hence, when the number of views increases, the number of contrastive loss terms also grows quadratically. As a result, in terms of either effectiveness or efficiency, it is challenging for the model to capture the cooperative effect of multiple views, which would be beneficial to the performance but cannot be learned from isolated individual views.

To address the above limitations, we propose a novel \textbf{Multi}-view \textbf{C}ontrastive learning framework for \textbf{B}undle \textbf{R}ecommendation, short as \textbf{MultiCBR}. 
First, we reorganize the multiple heterogeneous relations into three views: user-bundle interaction view, user-item interaction view, and bundle-item affiliation view, each of which focuses on one relation of UB, UI, and BI, as shown in Figure~\ref{fig:motivation}. This multi-view formulation ensures that all three graphs are fully exploited via bi-directional graph propagation, especially for the BI graph, which is sub-optimally modeled by previous works. At the same time, the three-view decomposition can better differentiate each type of relation. 
Second, to leverage the powerful contrastive learning into the \textit{multi-view} framework, we adopt a converse strategy as opposed to CrossCBR, \ie "early fusion and late contrast". As shown in Figure~\ref{fig:motivation}, we first fuse the multi-view representations into a unified one, and then apply self-supervised contrastive learning. Such a simple revision brings in two major advantages: 1) "early fusion" inherently captures user preference from both ego-view and cross-view (more details are presented in Section~\ref{subsec:fusion_prediction}); and 2) "early fusion and late contrast" eliminates the quadratic number of cross-view contrastive losses and just requires only two contrastive losses. This enables the incorporation of multiple views with few extra computational costs or optimization difficulties.

To evaluate MultiCBR, we conduct extensive experiments on three public benchmark datasets, \ie Youshu, NetEase, and iFashion. Overall speaking, MultiCBR outperforms SOTA methods (\eg CrossCBR~\cite{CrossCBR2022}) on all the three datasets. In particular on iFashion, we achieve over 20\% relative improvements. Further ablation and model studies verify the effectiveness of our model as well as the hypothesis. The key contributions of this work are summarized as follows:
\begin{itemize}[leftmargin=*]
    \item To the best of our knowledge, we are the first to propose a multi-view contrastive learning framework for bundle recommendation to fully exploit all the relations among users, bundles and items, especially addressing the BI sparsity problem by introducing the BI view.
    \item We propose the strategy of "early fusion and late contrast" to better capture user preference from both ego-view and cross-view, while introducing few extra expenses. 
    \item Extensive experiments on three benchmark datasets indicate that our method outperforms SOTA baselines. 
\end{itemize}
\section{Methodology} \label{sec:methodology}
\begin{figure*}
    \centering
    \includegraphics[width = 0.95\linewidth]{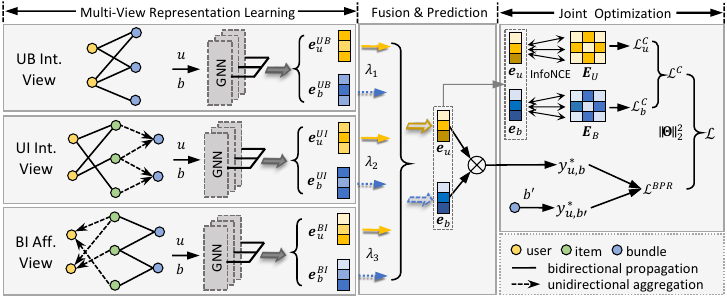}
    \caption{The overall framework of MultiCBR consists of three parts: 1) multi-view representation learning, including user-bundle interaction view (UB Int. View), user-item interaction view (UI Int. View), and bundle-item affiliation view (BI Aff. View); 2) multi-view fusion and prediction; and 3) joint optimization of BPR loss and self-supervised contrastive loss.}
    \vspace{-0.1in}
    \label{fig:framework}
\end{figure*}

In this section, we present the overall framework of our proposed method MultiCBR, as shown in Figure~\ref{fig:framework}. In particular, we first give a brief introduction of the problem formulation, followed by the three key modules of MultiCBR: 1) multi-view representation learning, 2) multi-view fusion and prediction, and 3) joint optimization.

\subsection{Problem Formulation} \label{subsec:problem_formulation}
The problem of bundle recommendation aims to route a set of users $\mathcal{U}=\{u_1, u_2, \cdots, u_M\}$ to a set of pre-defined bundles $\mathcal{B}=\{b_1, b_2, \cdots, b_N\}$, where each bundle is composed from a set of items $\mathcal{I}=\{i_1, i_2, \cdots, i_O\}$. Even though various information can be used to train a bundle recommender system, in this paper, we are particularly interested in using the user-bundle interactions $\mathbf{X}_{M \times N}=\{x_{ub}|u\in{\mathcal{U}},b\in{\mathcal{B}}\}$, user-item interactions $\mathbf{Y}_{M \times O}=\{y_{ui}|u\in{\mathcal{U}},i\in{\mathcal{I}}\}$, and bundle-item affiliation information $\mathbf{Z}_{N \times O}=\{z_{bi}|b\in{\mathcal{B}},i\in{\mathcal{I}}\}$, where $M,N,O$ are the number of users, bundles, and items, respectively. Given the three binary-valued matrix $\mathbf{X}$, $\mathbf{Y}$, and $\mathbf{Z}$, where $\mathbf{X}$ and $\mathbf{Y}$ are the historical interaction data, the target is to predict unseen user-bundle interactions. 

\subsection{Multi-view Representation Learning} \label{subsec:rep_learning}
In this work, we devise a multi-view representation learning framework that consists of three views: user-bundle interaction view, user-item interaction view, and bundle-item affiliation view, each of which specifically concentrates on one type of relations and learns view-tailored user and bundle representations. We separately learn user and bundle representations from three views, instead of blending the relations together (\eg building a tripartite graph~\cite{BundleNet2020}), since the mixture of relations disturbs each other and results in poor overall performance~\cite{CrossCBR2022}. The representation learning of the three views is as follows.

\subsubsection{User-Bundle Interaction View}
In this view, we aim to learn user and bundle representations that capture user preference based on the user-bundle CF signals. Based on the user-bundle interaction matrix $\mathbf{X}$, we construct a user-bundle bipartite graph (UB graph) and utilize the SOTA GNN-based CF signal modeling model LightGCN~\cite{LightGCN2020} to learn the representations of users and bundles. In particular, the graph propagation is depicted as:
\begin{equation} \label{eq_1}
\left\{
\begin{aligned}
    \mathbf{e}_{u}^{UB(k)} &= \sum_{b \in \mathcal{N}_u^{UB}}{\frac{1}{\sqrt{|\mathcal{N}_u^{UB}|}\sqrt{|\mathcal{N}_b^{UB}|}}\mathbf{e}^{UB(k-1)}_{b}}, \\
    \mathbf{e}_{b}^{UB(k)} &= \sum_{u \in \mathcal{N}_b^{UB}}{\frac{1}{\sqrt{|\mathcal{N}_b^{UB}|}\sqrt{|\mathcal{N}_u^{UB}|}}\mathbf{e}^{UB(k-1)}_{u}},
\end{aligned}
\right.
\end{equation}
where $\mathbf{e}_{u}^{UB(k)}, \mathbf{e}_{b}^{UB(k)} \in \mathbb{R}^{d}$ are the $k$-th layer's embeddings for user $u$ and bundle $b$; $d$ is the embedding dimension; $\mathcal{N}_u^{UB}$ and $\mathcal{N}_b^{UB}$ are the neighbors of the user $u$ and bundle $b$ in the UB graph. We use superscript $UB$ to denote the representations are for the user-bundle interaction view. After obtaining embeddings from different layers, we pool them into unified bundle-level representations:
\begin{equation} \label{eq_2}
    \mathbf{e}^{UB}_{u} = \frac{1}{K} \sum^K_{k=0}{\mathbf{e}^{UB(k)}_{u}}, \ \ \ \ \mathbf{e}^{UB}_b = \frac{1}{K} \sum^K_{k=0}{\mathbf{e}^{UB(k)}_b}.
\end{equation}

\subsubsection{User-Item Interaction View}
The user-item interaction view targets at modeling the user-item CF signals and generate corresponding user and bundle representations. In particular, we first employ a graph learning module over the user-item interaction graph to obtain the user and item representations. It not only captures the user preference over individual items but yields the item embeddings serving to represent the bundle at a fine-grained level. In order to yield bundle representations in this view, we perform a uni-directional aggregation from the items to their corresponding bundles, which is guided by the bundle-item affiliation graph. 

Based on the user-item interaction matrix $\mathbf{Y}$, we can build a bipartite UI graph. Analogous to the bundle level representation learning, a LightGCN is used to model the user-item CF signals. We present the details of graph propagation as follows:
\begin{equation} \label{eq_3}
\left\{
\begin{aligned}
    \mathbf{e}_{u}^{UI(k)} &= \sum_{i \in \mathcal{N}_u^{UI}}{\frac{1}{\sqrt{|\mathcal{N}_u^{UI}|}\sqrt{|\mathcal
    {N}_i^{UI}|}}\mathbf{e}^{UI(k-1)}_i}, \\
    \mathbf{e}_{i}^{UI(k)} &= \sum_{u \in \mathcal{N}_i^{UI}}{\frac{1}{\sqrt{|\mathcal{N}_i^{UI}|}\sqrt{|\mathcal{N}_u^{UI}|}}\mathbf{e}^{UI(k-1)}_{u}},
\end{aligned}
\right.
\end{equation}
where $\mathbf{e}_{u}^{UI(k)}, \mathbf{e}_{i}^{UI(k)} \in \mathbb{R}^{d}$ are embeddings for user $u$ and item $i$ at the $k$-th layer; $\mathcal{N}_u^{UI}$ and $\mathcal{N}_i^{UI}$ are the neighbors of the user $u$ and item $i$ in the UI graph. The superscript $UI$ indicates that the representations are used for the user-item interaction view. Similar to the UB graph learning, the layer pooling is conducted to obtain the item-level representations $\mathbf{e}^{UI}_{u}$ and $\mathbf{e}^{UI}_i$:
\begin{equation} \label{eq_4}
    \mathbf{e}^{UI}_{u} = \frac{1}{K} \sum^K_{k=0}{\mathbf{e}^{UI(k)}_{u}}, \ \ \ \ \mathbf{e}^{UI}_{i} = \frac{1}{K} \sum^K_{k=0}{\mathbf{e}^{UI(k)}_{i}}.
\end{equation}
We aggregate the items representations and devise the bundle representation as follows:
\begin{equation} \label{eq_5}
    \mathbf{e}^{UI}_{b} = \frac{1}{|\mathcal{N}_b^{BI}|}\sum_{i \in \mathcal{N}_b^{BI}}\mathbf{e}^{UI}_i.
\end{equation}
where $\mathcal{N}_b^{BI}$ is the first-order neighbors of bundle $b$ in the BI graph. Even though we utilize both UI and BI graphs in this view, only UI CF signals are well captured via bi-directional graph learning, while the BI graph is just used to pool the item embeddings into bundle representation and less explored. Therefore, we incorporate a third view to specifically mine the BI graph patterns, presented in the following section.

\subsubsection{Bundle-Item Affiliation View}
The bundle-item affiliation view concentrates on the bundle composition information, which should be injected into both user and bundle representations. We conduct bi-directional information propagation over the bundle-item affiliation graph while employ a uni-directional information aggregation from items to users over the user-item interaction graph. To be noted, this view does not incorporate any additional data compared with the user-item interaction view, nevertheless, it can alter the concentration from user-item CF patterns to the bundle-item composition patterns. Therefore, the bundle composition information will affect both user and bundle representations in this view, endowing the overall model with more information from bundle composition patterns.

Based on the bundle-item affiliation matrix $\mathbf{Z}$, we curate a bipartite BI graph, and then similarly adopt a LightGCN kernel to learn on the graph as follows:
\begin{equation} \label{eq_6} 
\left\{
\begin{aligned}
    \mathbf{e}_{b}^{BI(k)} &= \sum_{i \in \mathcal{N}_b^{BI}}{\frac{1}{\sqrt{|\mathcal{N}_b^{BI}|}\sqrt{|\mathcal
    {N}_i^{BI}|}}\mathbf{e}^{BI(k-1)}_i}, \\
    \mathbf{e}_{i}^{BI(k)} &= \sum_{b \in \mathcal{N}_i^{BI}}{\frac{1}{\sqrt{|\mathcal{N}_i^{BI}|}\sqrt{|\mathcal{N}_b^{BI}|}}\mathbf{e}^{BI(k-1)}_{b}},
\end{aligned}
\right.
\end{equation}
where $\mathbf{e}_{b}^{BI(k)}, \mathbf{e}_{i}^{BI(k)} \in \mathbb{R}^{d}$ are the $k$-th layer's embeddings for bundle $b$ and item $i$; $\mathcal{N}_b$ and $\mathcal{N}_i$ are the neighbors of the bundle $b$ and item $i$ in the BI graph. The superscript $BI$ represents that the representations are for the bundle-item affiliation view. Similar to Equation~\ref{eq_2} and Equation~\ref{eq_4}, we aggregate the embeddings from different layers and obtain the representations $\mathbf{e}^{BI}_{b}$ and $\mathbf{e}^{BI}_i$, defined as:
\begin{equation} \label{eq_7}
    \mathbf{e}^{BI}_{b} = \frac{1}{K} \sum^K_{k=0}{\mathbf{e}^{BI(k)}_{b}}, \ \ \ \ \mathbf{e}^{BI}_{i} = \frac{1}{K} \sum^K_{k=0}{\mathbf{e}^{BI(k)}_{i}}.
\end{equation}
To obtain the user representation according to the bundle-item compositional representations, we aggregate the representations of items that each user interacts with, represented as:
\begin{equation} \label{eq_8}
    \mathbf{e}^{BI}_{u} = \frac{1}{|\mathcal{N}_u^{UI}|}\sum_{i \in \mathcal{N}_u^{UI}}\mathbf{e}^{BI}_i.
\end{equation}
where $\mathcal{N}_u^{UI}$ are the items that have been interacted with user $u$ in the UI graph.

\subsection{Multi-view Fusion and Prediction} \label{subsec:fusion_prediction}
By performing the graph convolutional operations on three view-specific graphs (\ie UB, UI, and BI graphs), three representations are obtained for each user and bundle, \ie $\{\mathbf{e}^{UB}_u, \mathbf{e}^{UI}_u, \mathbf{e}^{BI}_u\}$ and $\{\mathbf{e}^{UB}_b, \mathbf{e}^{UI}_b, \mathbf{e}^{BI}_b\}$. In this section, we introduce a simple strategy to fuse the multi-view representations with a set of view coefficients and yield the final representation for prediction. More importantly, we deem that the inner-product prediction function following an early fused representations is able to capture both ego-view and cross-view user preference. Therefore, the explicit cross-view alignment proposed by CrossCBR~\cite{CrossCBR2022}, which is cumbersome for computation and optimization, is no longer required.

\subsubsection{Multi-view Fusion} We aim to fuse the heterogeneous views into a unified one in the representation level. Previous approaches~\cite{BGCN2020,CrossCBR2022} just combine the predictions of two views (as shown in Figure~\ref{fig:motivation}), which neglect the feature-level cooperation. Moreover, they just use equal weights to combine multiple views, which is sub-optimal for cross-view cooperation since each view contributes differently to the overall preference modeling. Hence, in order to capture the heterogeneity of multiple views and achieve better cooperative effects across multiple views, we early fuse the multi-view representations with a set of \textit{view coefficients}, thus obtain the final user and bundle representations $\mathbf{e}_{u}$ and $\mathbf{e}_{b}$:
\begin{equation} \label{eq_9}
\left\{
\begin{aligned}
    \mathbf{e}_{u} = \lambda_{1}\mathbf{e}^{UB}_u + \lambda_{2}\mathbf{e}^{UI}_u + \lambda_{3}\mathbf{e}^{BI}_u, \\
    \mathbf{e}_{b} = \lambda_{1}\mathbf{e}^{UB}_b + \lambda_{2}\mathbf{e}^{UI}_b + \lambda_{3}\mathbf{e}^{BI}_b,
\end{aligned}
\right.
\end{equation}
where $\{\lambda_{1}, \lambda_{2}, \lambda_{3}\}$ are the view coefficients to balance the contributions of three views and $\lambda_{1}+\lambda_{2}+\lambda_{3}=1$. 

\subsubsection{Prediction} We generate the preference score $y^*_{u,b}$ for user $u$ on bundle $b$ by applying the inner-product between the overall user and bundle representations, denoted as $y^*_{u,b}=\mathbf{e}_u \cdot \mathbf{e}_b$. Although it is common to employ inner-product as the preference scoring function in recommender systems, we argue that the inner-product on the fused representations inherently models both the cross-view and ego-view preference. By replacing $\mathbf{e}_{u}, \mathbf{e}_{b}$ with Equation~\ref{eq_9}\footnote{The view coefficients $\{\lambda_{1},\lambda_{2},\lambda_{3}\}$ are omitted for simplicity.}, we can rewrite $y^*_{u,b}$ and group the terms into two types of preference: \textit{cross-view preference} and \textit{ego-view preference}, denoted as:
\begin{equation} \label{eq_10}
\begin{aligned}
  y^*_{u,b} &= (\mathbf{e}^{UB}_u + \mathbf{e}^{UI}_u + \mathbf{e}^{BI}_u) \cdot (\mathbf{e}^{UB}_b + \mathbf{e}^{UI}_b + \mathbf{e}^{BI}_b) \\
      &= 
      \!\begin{aligned}[t]
      &\underbrace{\bigl(\mathbf{e}^{UB}_{u} \cdot (\mathbf{e}^{UI}_b + \mathbf{e}^{BI}_b) + \mathbf{e}^{UI}_{u} \cdot (\mathbf{e}^{UB}_b + \mathbf{e}^{BI}_b) + \mathbf{e}^{BI}_{u} \cdot (\mathbf{e}^{UB}_b + \mathbf{e}^{UI}_b)\bigr)}_{\text{cross-view preference}} \\
      &+\underbrace{(\mathbf{e}^{UB}_{u} \cdot \mathbf{e}^{UB}_b+\mathbf{e}^{UI}_{u} \cdot \mathbf{e}^{UI}_b+\mathbf{e}^{BI}_{u} \cdot \mathbf{e}^{BI}_b)}_{\text{ego-view preference}}
      \end{aligned}
\end{aligned}
\end{equation}
where the cross-view preference is the inner-product of the user and bundle representations from different views, while the ego-view preference is the inner-product of the user and bundle representations from the same view. Contemporary methods~\cite{BGCN2020,CrossCBR2022} follow a late-fusion strategy, in which they separately calculate the preference for each view and then sum the prediction scores. In other words, they just model the ego-view preference. However, our early fusion strategy inherently enable the model to capture both ego-view and cross-view preference. To be noted, CrossCBR can also achieve cross-view cooperation through the cross-view contrastive learning. However, it just increases the cosine-similarity of the representations from different views. We argue that our way of cross-view preference modeling is more direct and effective. 

\subsection{Joint Optimization} \label{subsec:joint_optimization}
We employ a joint optimization protocol by combining the self-supervised contrastive loss with the typical BPR (Bayesian Personalized Ranking~\cite{BPR2012}) loss. 

\subsubsection{Contrastive Loss} Contrastive learning has witnessed great success in computer vision~\cite{SimCLR2020,SimSiam2021}, natural language processing~\cite{SimSiam2021}, and graph learning~\cite{DGI2019,GMI2020,MVGRL2020,GRACE2020,GCA2021}. Slightly after that, recommender systems quickly adapt contrastive learning into various recommendation scenarios, including CF-based user-item~\cite{SGL2021,SimGCL2022}, sequential (session)~\cite{S32020}, cold-start~\cite{CLC4Rec2021}, multi-behavior~\cite{MMCLR2022}, cross-domain recommendations~\cite{CCDR2022}, and \textit{etc}. CrossCBR~\cite{CrossCBR2022} and MIDGN~\cite{MIDGN2022} are the first to incorporate contrastive learning into bundle recommendation. CrossCBR constructs cross-view contrastive losses to enhance the representation affinity of the same user (bundle) from different views, while decrease that of different users (bundles). Analogously, MIDGN builds contrastive losses between global and local views, where different views of the same user (bundle) form positive pairs. The main idea behind these two works is to use cross-view contrastive loss to achieve cross-view cooperation.

In this paper, we discard the explicit cross-view contrastive loss and just adopt a simple self-supervised contrastive loss on the unified representations. Since the "early fusion" strategy already endows MultiCBR with the capability of multi-view cooperation modeling, it is no longer necessary to build cross-view contrastive losses, and the self-supervised contrastive loss is sufficient to enhance the model with capabilities such as countering sparsity or noise reported in previous works~\cite{SGL2021,SimGCL2022}. 

To construct the positive pairs for applying contrastive learning and also enhance the robustness to counter potential noise, we adopt data augmentation to generate different representations of the same user or bundle. In detail, for each user $u$, we obtain two different representations $\mathbf{e}_{u}^{\prime}$ and $\mathbf{e}_{u}^{\prime\prime}$ indicating the same user $u$ under different data augmentations. $\mathbf{e}_{u}^{\prime}$ and $\mathbf{e}_{u}^{\prime\prime}$ are considered as a positive user pair. Similarly, for each bundle $b$, $\mathbf{e}_{b}^{\prime}$ and $\mathbf{e}_{b}^{\prime\prime}$ are generated as the positive bundle pair. For data augmentation, we follow the previous works \cite{SGL2021, CrossCBR2022, SimGCL2022, yu2022xsimgcl} and adopt three different methods: edge dropout (ED), message dropout (MD) and noise augmentation (Noise). The edge dropout is based on the graph structure, where we randomly drop a small percentage of edges in the input graph to generate an augmented graph. Different representations are obtained by learning on augmented graph structures. Both message dropout and noise augmentation are embedding-based. Message dropout randomly masks some elements of the propagated embeddings with a certain dropout ratio $\rho$. Noise augmentation directly adds a small noise vector $\Delta$ which subjects to $||\Delta||^2 = \epsilon$ and $\epsilon$ is a small constant. Stronger data augmentation methods could further enhance the model performance and those data augmentation techniques developed in general self-supervised graph contrastive learning can be easily adapted to our framework without any specialized consideration, which is left for future work to explore. 

For contrastive loss, we adopt InfoNCE~\cite{InfoNCE2010} built upon the generated contrastive pairs ($\mathbf{e}_{u}^{\prime}$,  $\mathbf{e}_{u}^{\prime\prime}$) and ($\mathbf{e}_{b}^{\prime}$,  $\mathbf{e}_{b}^{\prime\prime}$). The equations are as follows:
\begin{equation} \label{eq:eq_11}
    \left\{
        \begin{aligned}
            \mathcal{L}^C_{u} &= \frac{1}{|\mathcal{U}|}\sum_{u \in \mathcal{U}}{-\text{log}\frac
            {\text{exp}({\text{cos}(\mathbf{e}_{u}^{\prime}, \mathbf{e}_{u}^{\prime\prime})/\tau})}
            {\sum_{v \in \mathcal{U}}{\text{exp}({\text{cos}(\mathbf{e}_{u}^{\prime}, \mathbf{e}_{v}^{\prime\prime})/\tau})}}}, \\
            \mathcal{L}^C_{b} &= \frac{1}{|\mathcal{B}|}\sum_{b \in \mathcal{B}}{-\text{log}\frac
            {\text{exp}({\text{cos}(\mathbf{e}_{b}^{\prime}, \mathbf{e}_{b}^{\prime\prime})/\tau})}
            {\sum_{p \in \mathcal{B}}{\text{exp}({\text{cos}(\mathbf{e}_{b}^{\prime}, \mathbf{e}_{p}^{\prime\prime})/\tau})}}},
        \end{aligned}
    \right.
\end{equation}
where $\text{cos}(,)$ is the cosine similarity function, $\tau$ is the hyper-parameter known as \textit{temperature} softmax. We adopt the popular implementation of in-batch negative sampling~\cite{SGL2021} to construct the negative pairs. 

To be noted, we only need two contrastive loss terms no matter how many views are included. Meanwhile, for previous cross-view contrastive learning framework, when the number of views increases (\eg the introduction of a third view in our work), the cross-view contrastive loss terms grow quadratically, causing extra computational and optimization overhead. In summary, MultiCBR not only retains the advantage of contrastive learning but also reduces the computational and optimization overhead.

\subsubsection{Optimization} BPR Loss has been widely used as the main learning objective for current bundle recommendation models, formally obtained by:
\begin{equation} \label{eq:bpr_loss}
    \mathcal{L}^{BPR} = \sum_{(u,b,b^{\prime}) \in Q}{-\text{ln} \sigma (y^*_{u,b} - y^*_{u,b^{\prime}})},
\end{equation}
where $b^{\prime}$ is the negative sample that are randomly chosen from the bundles that have not been interacted with user $u$, $Q = \{(u,b,b^{\prime})|u \in \mathcal{U}, b, b^{\prime} \in \mathcal{B}, x_{ub}=1, x_{ub^{\prime}}=0 \}$, and $\sigma(\cdot)$ is the sigmoid function. Finally, we optimize the whole framework by the joint loss:
\begin{equation}
    \mathcal{L} = \mathcal{L}^{BPR} + \beta_{1}\mathcal{L}^C + \beta_{2}{\Vert \mathbf{\Theta} \rVert}_2^2,
\end{equation}
where $\mathcal{L}^C=\frac{1}{2}(\mathcal{L}^C_{u}+\mathcal{L}^C_{b})$, $\beta_1$ and $\beta_2$ are hyper-parameters to balance different loss terms, and ${\Vert \mathbf{\Theta} \rVert}_2^2$ is the L2 regularization term. To be noted, same as CrossCBR, our model also just requires three sets of embeddings for users, bundles, and items, denoted as $\mathbf{\Theta}=\{\mathbf{E}_U^{(0)}, \mathbf{E}_B^{(0)}, \mathbf{E}_I^{(0)}\}$.

\subsection{Complexity Analysis} \label{subsec:complexity_analysis}
For space complexity, the parameters of MultiCBR only include three sets of embeddings: $\mathbf{E}_U^{(0)}$, $\mathbf{E}_B^{(0)}$, $\mathbf{E}_I^{(0)}$. The total space complexity of MultiCBR is $\mathcal{O}((M+N+O)d)$. Compared with CrossCBR, MultiCBR introduces the additional bundle-item affiliation view without adding extra parameters.

For time complexity, the computational cost of MultiCBR mainly comes from graph learning of three views and contrastive learning. The time complexity of graph learning in MultiCBR is $O((2K|E_{UB}| + (2K+1)(|E_{UI}| + |E_{BI}|))ds\frac{|E_{UB}|}{T})$, where $|E_{UB}|$, $|E_{UI}|$, $|E_{BI}|$ are the number of edges in UB graph, UI graph, BI graph respectively, $K$ is the number of propagation layers, $d$ is the embedding size, $s$ the number of epochs and $T$ is the batch size. In comparison, the time complexity of graph learning in CorssCBR is $O((2K|E_{UB}| + 2K|E_{UI}| + |E_{BI}|)ds\frac{|E|}{T})$. The extra time complexity in MultiCBR caused by introducing the BI view is $O((|E_{UI}| + 2K|E_{BI}|)ds\frac{|E|}{T})$. The time complexity for calculating contrastive loss is $O(2d(T + 1)s|E|)$, which is the same as CrossCBR. Owing to the "early fusion and late contrast" structure, the time consumption of contrastive loss does not increase as the number of views grows.
\section{Experiments} \label{sec:experiment}
To evaluate our proposed approach, we conduct experiments on three public bundle recommendation datasets: Youshu, NetEase, and iFashion. In particular, we aim to answer the following research questions:

\begin{itemize}[leftmargin=*]
    \item \textbf{RQ1: } Can MultiCBR outperform the SOTA baseline models?
    \item \textbf{RQ2: } Are all the views and the "early fusion and late contrast" framework helpful for the overall performance?
    \item \textbf{RQ3: } What are the key characteristics of MultiCBR \wrt user preference modeling and efficiency? 
\end{itemize}

\subsection{Experimental Settings}
\textbf{Datsets}: We follow the three public bundle datasets from CrossCBR, \ie Youshu~\cite{DAM2019}, NetEase~\cite{cao2017embedding} and iFashion~\cite{POG2019,HFGN2020}, corresponding to scenarios of book list, music list and fashion outfit, respectively. We adopt the same split of training, validation, and testing with CrossCBR. And the dataset statistics are depicted in Table~\ref{tab:dataset}. NDCG@K and Recall@K are leveraged as the evaluation metrics, where $K \in \{20, 40\}$ and all-ranking protocol is used. 

\begin{table}[t]
\begin{center}
\caption{Dataset Statistics.}
\label{tab:dataset}
\resizebox{0.7\textwidth}{!}{
    \begin{tabular}{ccccccc}
        \hline
        Dataset & \#U & \#I & \#B & \#U-I & \#U-B & \#Avg.I/B \\
        \hline
        Youshu   & 8,039  & 32,770  & 4,771  & 138,515   & 51,377    & 37.03 \\
        NetEase  & 18,528 & 123,628 & 22,864 & 1,128,065 & 302,303   & 77.80 \\
        iFashion & 53,897 & 42,563 & 27,694 & 2,290,645 & 1,679,708 & 3.86 \\
        \hline
    \end{tabular}
}
\end{center}
\end{table}

\begin{table*}[t]
\caption{The overall performance comparison. R represents \textit{Recall} and N represents \textit{NDCG}.}
\label{tab:overall_performance}
\centering
\setlength{\tabcolsep}{1mm}{
    \resizebox{\textwidth}{!}{
        \begin{tabular}{l | cccc | cccc | cccc}
        \hline
        \multirow{2}{*}{Model} & \multicolumn{4}{c|}{Youshu} &\multicolumn{4}{c|}{NetEase} &\multicolumn{4}{c}{iFashion} \\
        \cline{2-13} & R@20 & N@20 & R@40 & N@40 & R@20 & N@20 & R@40 & N@40 & R@20 & N@20 & R@40 & N@40 \\
       \hline
        \hline
       \textbf{MFBPR}     & 0.1959	& 0.1117 & 0.2735 & 0.1320 & 0.0355 & 0.0181 & 0.0600 & 0.0246 & 0.0752 & 0.0542 & 0.1162 & 0.0687 \\
        \textbf{LightGCN}  & 0.2286 & 0.1344 & 0.3190 & 0.1592 & 0.0496 & 0.0254 & 0.0795 & 0.0334 & 0.0837 & 0.0612 & 0.1284 & 0.0770 \\ 
        \textbf{SGL} & 0.2622 & 0.1551 & 0.3630 & 0.1825 & 0.0716 & 0.0386 & 0.1104 & 0.0488 & 0.1000 & 0.0739 & 0.1498 & 0.0915 \\
        \textbf{XSimGCL} & 0.2644 & 0.1560 & 0.3569 & 0.1814 & 0.0710 & 0.0376 & 0.1079 & 0.0475 & 0.1003 & 0.0744 & 0.1501 & 0.0920\\
        \textbf{LightGCL} & 0.2710 & 0.1556 & 0.3693 & 0.1827 & 0.0705 & 0.0371 & 0.1122 & 0.0481 & 0.0894 & 0.0653 & 0.1362 & 0.0818\\
        \hline
        \hline
        \textbf{DAM} & 0.2082 & 0.1198 & 0.2890 & 0.1418 & 0.0411 & 0.0210 & 0.0690 & 0.0281 & 0.0629 & 0.0450 & 0.0995 & 0.0579 \\
        \textbf{BundleNet} & 0.1895 & 0.1125 & 0.2675 & 0.1335 & 0.0391 & 0.0201 & 0.0661 & 0.0271 & 0.0626 & 0.0447 & 0.0986 & 0.0574 \\
        \textbf{BGCN} & 0.2436 & 0.1329 & 0.3379 & 0.1589 & 0.0626 & 0.0328 & 0.0988 & 0.0424 & 0.0802 & 0.0582 & 0.1239 & 0.0736 \\     
        \textbf{MIDGN} & 0.2682 & 0.1527 & 0.3712 & 0.1808 & 0.0678 & 0.0343 & 0.1085 & 0.0451 & 0.0856 & 0.0473 & 0.1299 & 0.0593 \\
        \textbf{CrossCBR} & \underline{0.2823} & \underline{0.1670} & \underline{0.3787} & \underline{0.1939} & \underline{0.0844} & \underline{0.0458} & \underline{0.1255} & \underline{0.0567} & \underline{0.1132} & \underline{0.0872} & \underline{0.1623} & \underline{0.1045} \\
        \hline
        \hline
        \textbf{MultiCBR} & \textbf{0.2842} & \textbf{0.1693} & \textbf{0.3946} & \textbf{0.1994} & \textbf{0.0907} & \textbf{0.0494} & \textbf{0.1355} & \textbf{0.0612} & \textbf{0.1500} & \textbf{0.1203} & \textbf{0.2033} & \textbf{0.1390} \\
        \textbf{\%Improv.} & 0.66 & 1.39 & 4.20 & 2.87 & 7.49 & 7.82 & 8.02 & 7.86 & 32.43 & 37.96 & 25.26 & 33.09 \\
        \hline
        \end{tabular}
    }
}
\end{table*}

\subsubsection{Compared Methods}
Both user-item and bundle-specific recommender models are leveraged for baselines.

\noindent \textbf{The User-item Recommender Models} just utilize the user-bundle interaction data without considering the affiliated items within each bundle. The following methods are considered: 
\begin{itemize}[leftmargin=*]
\item \textbf{MFBPR}~\cite{BPR2012}: Matrix Factorization optimized by the BPR loss.
\item \textbf{LightGCN}~\cite{LightGCN2020}: it utilizes a light-version graph learning kernel to model the CF signals.
\item \textbf{SGL}~\cite{SGL2021}: it applies self-supervised contrastive learning to the LightGCN model and achieves SOTA performance.
\item \textbf{XSimGCL}~\cite{XSimGCL2022}: it also leverages self-supervised contrastive graph learning, while it novely proposes a add small-scale random noise as data augmentation.
\item \textbf{LightGCL}~\cite{LightGCL2023}: this method utilizes a novel method of SVD to generate augmented views for self-supervised contrastive learning.
\end{itemize}

\noindent \textbf{The Bundle-specific Recommender Models} take into account all the relations among users, bundles and items. We incorporate the following baselines: 
\begin{itemize}[leftmargin=*]
\item \textbf{DAM}~\cite{DAM2019}: it generates bundle representation from its included items through an attention score and employs multi-task learning to learn both user-item and user-bundle preferences.
\item \textbf{BundleNet}~\cite{BundleNet2020}: a tripartite graph among user, bundle and item is leveraged to learn the representations via GNN.
\item \textbf{BGCN}~\cite{BGCN2020}: it organizes the user preference into bundle and item views, each of which is modeled by a GCN module. It first makes predictions on each view and then sum the prediction scores.
\item \textbf{MIDGN}~\cite{MIDGN2022}: it formulates user preference in to local and global views and leverages the intent disentanglement to learn user's intents for bundle recommendation.
\item \textbf{CrossCBR}~\cite{CrossCBR2022}: it is the SOTA method, which utilizes a cross-view contrastive learning to achieve cross-view cooperative association based on the BGCN's view construction method.
\end{itemize}
Among all the baselines, \textbf{SGL}, \textbf{XSimGCL}, \textbf{LightGCL}, \textbf{MIDGN} and \textbf{CrossCBR} incorporate contrastive learning and are stronger than other non-contrast baselines.

\subsubsection{Hyper-parameter Settings}
We inherit most of the settings from CrossCBR~\cite{CrossCBR2022}, which uses embedding size 64, Xavier~\cite{Xavier2010} initialization, batch size 2048, and Adam optimizer~\cite{Adam2014}. For the hyper-parameters, we set the maximal number of layers $K=2$ and tune $\{\lambda_1, \lambda_2, \lambda_3\}$ with the range of $[0, 1]$. We tune $\tau$, $\beta_1$, $\beta_2$ with the range of $\{0.05, 0.1, 0.15, 0.2, 0.25, 0.3, 0.4\}$, $\{0.025, 0.05, 0.1, 0.15, 0.2, 0.3, 0.4\}$, $\{10^{-5}, 10^{-6}, 10^{-7}\}$. We re-implement several most representative baselines \ie SGL, XSimGCL, LightGCL, BGCN and CrossCBR, and directly take the reported results of other models from the original paper of CrossCBR and MIDGN, since we use the identical dataset and split. Since MIDGN does not take into account the iFashion dataset, we use their released code to obtain the result. For all the models mentioned above and our proposed MultiCBR, the model performance could vary in each training due to randomness, which is relatively more significant in small datasets like Youshu. For fair comparison, we repeated the model training three times with the same hyper-parameter and take the average performance as the reported value.

\subsection{Performance Comparison (RQ1)} \label{subsec:performance_comparison}
The overall performance is presented in Table~\ref{tab:overall_performance}. First, MultiCBR outperforms all the baseline methods on all three datasets. Especially on iFashion, we improve the SOTA performance by a large margin. This may because the bundle-item graph is very sparse in iFashion (there is only 3.86 items in each bundle as shown in Table~\ref{tab:dataset}) and the newly introduced BI view is significant to capture the valuable BI composition patterns. For similar reasons, the performance improvement on Youshu is relatively weaker. Bundles in Youshu generally have more items, and more importantly, most items in the bundle possess rich user-item interactions. As a result, introducing BI view, which is originally designed for countering the BI sparsity problem, helps less with the performance on Youshu. We will further discuss the effects of graph learning on BI view thoroughly in section~\ref{subsubsec:effectiveness_BI_view}. Second, the contrastive learning-based methods achieve better performance. For the user-item recommender models, SGL, XSimGCL, and LightGCL perform similar and are better than the methods without contrastive learning. Analogously, MIDGN and CrossCBR beat all the other bundle-specific methods. This phenomenon shows that contrastive learning is a powerful and general approach to boosting bundle recommendation. Third, CrossCBR is the strongest baseline, justifying that the two-view formulation (\aka bundle and item view) plus cross-view contrastive learning is better than the view formulation in MIDGN (\aka local and global view). Finally, MultiCBR outperforms CrossCBR, indicating the effectiveness of the three-view representation learning and the novel framework of "early fusion and late contrast".  

\subsection{Ablation Study (RQ2)} \label{subsec:ablation_study}

\begin{table}[t]
\begin{center}
\caption{Ablation study of the key components of MultiCBR.}
\label{tab:ablation_study}
    \resizebox{0.65\textwidth}{!}{
        \begin{tabular}{l | cc | cc}
            \hline
            \multirow{2}{*}{Model} & \multicolumn{2}{c|}{NetEase} &\multicolumn{2}{c}{iFashion} \\
            \cline{2-5}
             & Recall@20 & NDCG@20 & Recall@20 & NDCG@20  \\
            \hline
            \hline
            \textbf{SGL} & 0.0716 & 0.0386 & 0.1000 & 0.0739 \\
            \textbf{MultiCBR-UI} & 0.0871 & 0.0472 & 0.1357 & 0.1110 \\
            \textbf{MultiCBR-BI} & 0.0882 & 0.0474 & 0.1328 & 0.1044 \\
            \hline
            \hline
            \textbf{CrossCBR} & 0.0844 & 0.0458 & 0.1132 & 0.0872 \\
            \textbf{CrossCBR+BI} & 0.0789 & 0.0420 & 0.1293 & 0.1042  \\
            \hline
            \hline
            \textbf{CrossCBR-CL} & 0.0608 & 0.0320 & 0.0852 & 0.0626 \\
            \textbf{MultiCBR-CL} & 0.0498 & 0.0260 & 0.0671 & 0.0482 \\
            \hline
            \hline
            \textbf{MultiCBR} & 0.0907 & 0.0494 & 0.1500 & 0.1203  \\
            \hline
        \end{tabular}
    }
\end{center}
\vspace{-0.2in}
\end{table}

To further identify the contribution of each module of MultiCBR, we conduct a series of ablation study. Since the performance improvement of Youshu is marginal, the ablation study and the model study (in the following subsection) are mainly conducted on the NetEase and iFashion datasets. First, in order to verify the effectiveness of each view, we individually remove two views of UI and BI and contruct two ablated models MultiCBR-UI and MultiCBR-BI. According to the results in Table~\ref{tab:ablation_study}, by removing either view of UI and BI, the performance drops. Nevertheless, MultiCBR-UI and MultiCBR-BI are still much better than SGL, which only has the UB view. We can conclude that both UI and BI graphs provide valuable information for the user-bundle preference modeling. 

Second, we aim to investigate whether the "early fusion and late contrast" framework is useful. Even though it is cumbersome to generalize CrossCBR from two views to multiple views, it is still doable to adapt the pair-wise cross-view contrastive loss to three views. In particular, we add the BI view to CrossCBR and build an extension, named CrossCBR+BI, which includes six cross-view contrastive losses for user and bundle on three pairs of views: UB-UI, UB-BI and UI-BI. To be noted, CrossCBR+BI is a faithful implementation of the "early contrast and late fusion" strategy that is adopted by CrossCBR, instead of purely incorporating an additional BI view. From the results in Table~\ref{tab:ablation_study}, we have the following observations. First, CrossCBR+BI outperforms CrossCBR on iFashion but under-perform CrossCBR on NetEase, implying that the modeling of multiple views is more challenging than two views, where the design for effective multi-view cooperation is the crux. The cross-view contrastive learning framework may fail to capture the cooperation across multiple views or even worsen the performance by the introduction of the BI view. Second, MultiCBR beats CrossCBR+BI on both datasets, showing that the "early fusion and late contrast" framework is more effective than the "early contrast and late fusion" framework in exploiting the multi-view inputs.

Third, to quantitatively validate the effectiveness of contrastive loss, we remove it from our model and build a variation MultiCBR-CL. Comparing with MultiCBR, MultiCBR-CL dramatically decreases the performance and is even worse than CrossCBR-CL (the model variation that is constructed by removing the contrastive loss module from CrossCBR). That is to say, "late contrast" is a prerequisite for the success of "early fusion" and the specific ordered combination is the key to MultiCBR. 

\subsection{Model Study (RQ3)} \label{subsec:model_study}
In this section, we aim to investigate the key characteristics of MultiCBR: 1) how does the auxiliary BI view help to improve the model performance, especially in terms of the BI sparsity issue? 2) how does MultiCBR capture the cross-view and ego-view preference under the "early fusion and late contrast" framework? 3) how do the key hyper-parameters, \ie the view coefficients, the contrastive loss weight, and the temperature in contrastive loss, affect the model performance? 4) how about the computational efficiency of MultiCBR?

\subsubsection{Effects of the BI view} \label{subsubsec:effectiveness_BI_view}
Due to the lack of graph propagation on the BI view, the performance of CrossCBR is largely restricted by the sparsity of the bundle-item graph, especially in iFashion dataset where the average bundle size is only $3.86$. Additionally, some items in the bundles never interact with any user. The embeddings of such items cannot be learned without the BI view. MultiCBR alleviates such weakness and enhances the recommendation performance through learning from all UB, UI and BI views. 

\begin{table}[t]
\begin{center}
\caption{Performance comparison on the Youshu dataset with sparsified bundle-item graph of various dropout rate.}
\label{tab:youshu_sparse}
    \resizebox{0.95\textwidth}{!}{
        \begin{tabular}{l | cc | cc | cc | cc | cc}
            \hline
            \multirow{2}{*}{Model} & \multicolumn{2}{c|}{No Drop} & \multicolumn{2}{c|}{Drop 10\%} &\multicolumn{2}{c|}{Drop 30\%} &\multicolumn{2}{c|}{Drop 50\%} &\multicolumn{2}{c}{Drop 80\%} \\
            \cline{2-11}
             & R@20 & N@20 & R@20 & N@20 & R@20 & N@20 & R@20 & N@20 & R@20 & N@20 \\
            \hline
            \hline
            \textbf{CrossCBR} & 0.2823 & 0.1670 & 0.2683 & 0.1582 & 0.2672 & 0.1563 & 0.2644 & 0.1546 & 0.2553 & 0.1477 \\
            \hline
            \hline
            \textbf{MultiCBR} & 0.2842 & 0.1693 & 0.2815 & 0.1677 & 0.2817 & 0.1678 & 0.2811 & 0.1675 & 0.2817 & 0.1672 \\
            \textbf{\%Improv.} & 0.67 & 1.38 & 4.89 & 6.03 & 5.43 & 7.35 & 6.32 & 8.35 & 10.33 & 13.22 \\
            \hline
        \end{tabular}
    }

\end{center}
\end{table}

\begin{table}[t]
\begin{center}
\caption{Performance comparison on the NetEase dataset with sparsified bundle-item graph of various dropout rate.}
\label{tab:netease_sparse}
    \resizebox{0.95\textwidth}{!}{
        \begin{tabular}{l | cc | cc | cc | cc | cc}
            \hline
            \multirow{2}{*}{Model} & \multicolumn{2}{c|}{No Drop} & \multicolumn{2}{c|}{Drop 10\%} &\multicolumn{2}{c|}{Drop 30\%} &\multicolumn{2}{c|}{Drop 50\%} &\multicolumn{2}{c}{Drop 80\%} \\
            \cline{2-11}
             & R@20 & N@20 & R@20 & N@20 & R@20 & N@20 & R@20 & N@20 & R@20 & N@20 \\
            \hline
            \hline
            \textbf{CrossCBR} & 0.0844 & 0.0458 & 0.0838 & 0.0450 & 0.0831 & 0.0446 & 0.0811 & 0.0436 & 0.0747 & 0.0398 \\
            \hline
            \hline
            \textbf{MultiCBR} & 0.0907 & 0.0494 & 0.0889 & 0.0481 & 0.0879 & 0.0477 & 0.0867 & 0.0470 & 0.0839 & 0.0456 \\
            \textbf{\%Improv.} & 7.46 & 7.86 & 5.99 & 6.97\% & 5.88 & 6.90 & 6.91 & 7.84 & 12.33 & 14.65 \\
            \hline
        \end{tabular}
    }

\end{center}
\end{table}

\textbf{B-I Sparsity Issue.} Among the three datasets, the performance improvement on iFashion is the largest because we think MultiCBR can address the BI sparsity issue, where the BI connection in iFashion is the most sparse one. However, according to the dataset statistics in Table~\ref{tab:dataset}, Youshu and NetEase are relatively dense in terms of the BI affiliation graph since each bundle has much more items than iFashion. A natural question raises: does MultiCBR still perform well on sparse-version of Youshu and NetEase? To verify this hypothesis, we randomly drop certain rate of BI connections in the Youshu and NetEase, generating multiple sparse datasets with different level of BI sparsity. The results are shown in Table~\ref{tab:youshu_sparse} and Table~\ref{tab:netease_sparse}. On Youshu dataset, the relative performance improvements on sparse datasets are more significant than the original dataset (\aka No Drop), demonstrating that MultiCBR will perform much better if the BI connection is sparser. On both Youshu and NetEase datasets, the overall performance of both CrossCBR and MultiCBR generally decreases when the drop ratio increases, showing that the sparser the BI graph the more challenging of the task. However, the relative performance improvements of MultiCBR compared with CrossCBR generally increases when the sparsity ratio grows, showing that MultiCBR is better than CrossCBR in tackling the BI sparsity issue. 

\begin{figure}[t]
    \centering
    \includegraphics[width = 0.9\linewidth]{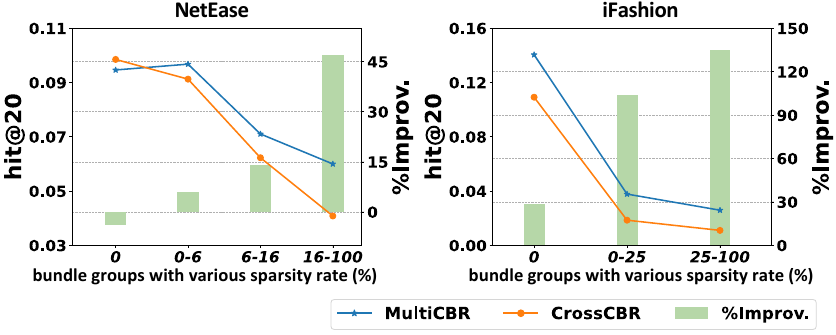}
    \caption{The performance comparison among bundle groups with different B-I-U sparsity rates.}
    \label{fig:model_study_BI_view}
\end{figure}

\textbf{B-I-U Sparsity Issue}. MultiCBR can alleviate the sparsity issue also because that it is able to learn better representations of items without user-item interactions. To verify this hypothesis, we split the bundles in iFashion and NetEase datasets into several groups according to the percentage of items without user interactions in each bundle (denoted as sparsity rate in Figure~\ref{fig:model_study_BI_view}). We call this as the B-I-U sparsity issue. The hit rates of the top 20 recommendation results (denoted as hit@20) of MultiCBR and CrossCBR are evaluated on each group of bundles. The results are shown in Figure~\ref{fig:model_study_BI_view}, where the horizontal axis implies the range of the sparsity rate of bundles in the same group. The lines in Figure~\ref{fig:model_study_BI_view} represent the evaluated model performance of CrossCBR and MultiCBR, and the bars measure the relative improvement of MultiCBR in each group. As shown in Figure~\ref{fig:model_study_BI_view}, when the percentage of items without user interactions in the bundle increases, the performance of both CrossCBR and MultiCBR decreases, implying the difficulty of representation learning for items without user interactions. Regarding the relative improvement of MultiCBR, in the first bundle group of NetEase, where all items in the bundle possess at least one user interaction, CrossCBR even slightly outperformed MultiCBR. But as the percentage of items without user interaction increases, MultiCBR soon outperforms CrossCBR, and the relative improvement grows rapidly. This phenomenon indicates that representation learning on BI view in MultiCBR effectively enhances the representation learning of items without user interactions, causing better overall performance. In iFashion dataset, MultiCBR outperforms CrossCBR significantly in all three bundles groups. We attribute this to the small average bundle size in iFashion dataset, where BI interactions are always sparse and graph learning on BI view greatly enhances the model performance. This further demonstrates the superiority of MultiCBR in modeling sparse BI signals.

\begin{figure}[t]
    \centering
    \includegraphics[width = 0.9\linewidth]{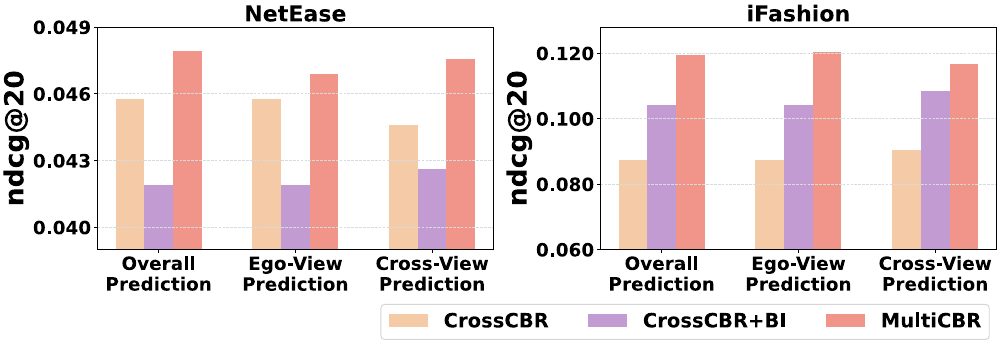}
    \caption{The cross-view and overall performance comparison for several ablated models.}
    \label{fig:model_study_1}
\end{figure}

\subsubsection{Cross- and ego-view preference modeling}
The key advantage of MultiCBR lies in that it can inherently model both the cross-view and ego-view preference, as presented in Equation~\ref{eq_10}. To justify this hypothesis, we separate the cross-view and ego-view terms of Equation~\ref{eq_10} and sum them up as prediction score for cross-view and ego-view, respectively. We rank all the candidate items according to the prediction scores and obtain the performance, which can directly reflect the corresponding preference modeling capacity. In Figure~\ref{fig:model_study_1}, we plot the cross-view, ego-view, and overall performance for three different models: CrossCBR, CrossCBR+BI, and MultiCBR. Overall speaking, MultiCBR performs best on the three types of prediction, indicating that our model is able to explicitly optimize both cross- and ego-view preference thus achieves the best overall performance. Comparing CrossCBR with MultiCBR on NetEase, the performance drop on cross-view prediction is larger than that on ego-view, showing that 
our method excels in modeling cross-view preference. Once again, CrossCBR+BI under-performs CrossCBR on NetEase for both ego-view and cross-view, showing that an effective framework may fail the multi-view cooperation.

\begin{table}[t]
\begin{center}
\renewcommand{\arraystretch}{1.3}
\caption{The cross-view alignment and dispersion analysis of the representations. For alignment, the larger the better, while for dispersion, the lower the better.}
\label{tab:alignment_dispersion}
    \resizebox{0.65\textwidth}{!}{
        \begin{tabular}{c | cc | cc}
            \hline
            \multirow{2}{*}{Metrics} & \multicolumn{2}{c|}{NetEase} &\multicolumn{2}{c}{iFashion} \\
            \cline{2-5}
             & CrossCBR+BI & MultiCBR & CrossCBR+BI & MultiCBR  \\
            \hline
            \hline
            $\mathbf{A}^U_{UB,UI}$ & 0.9218	& 0.9147 & 0.9276 & 0.7857 \\
            $\mathbf{A}^U_{UB,BI}$ & 0.7637	& 0.4473 & 0.8977 & 0.7606 \\
            $\mathbf{A}^U_{UI,BI}$ & 0.9106	& 0.5770 & 0.9728 & 0.9220 \\
            $\mathbf{A}^B_{UB,UI}$ & 0.5671	& 0.4116 & 0.9425 & 0.6409 \\
            $\mathbf{A}^B_{UB,BI}$ & 0.7921	& 0.7712 & 0.9556 & 0.8009 \\
            $\mathbf{A}^B_{UI,BI}$ & 0.8431	& 0.6941 & 0.9502 & 0.7614 \\
            \hline
            \hline
            $\mathbf{D}^U$ & 0.0136 & 0.0111 & 0.0006 & 0.0031 \\
            $\mathbf{D}^B$ & 0.0119	& 0.0162 & 0.0070 & 0.0099 \\
            \hline
        \end{tabular}
    }
\end{center}
\end{table}

We also analyze the Alignment-Dispersion~\cite{CrossCBR2022,SimCSE2021,AlignUniform2020} properties of the learned representations to characterize our model, as shown in Table~\ref{tab:alignment_dispersion}. In particular, we calculate the cosine similarity of representations of the same user (bundle) from different views to characterize the \textit{alignment} properties of the model, denoted as $\mathbf{A}^U$ and $\mathbf{A}^B$ for users and bundles respectively (where the subscript indicates the alignment pair, \eg $\mathbf{A}^U_{UB,UI}$ represents the user representation alignment between UB and UI view). We calculate the cosine similarity of the representations of different users (bundles) to characterize the \textit{dispersion} properties of the model, denoted as $\mathbf{D}^U$ and $\mathbf{D}^B$ for users and bundles, respectively (for CrossCBR+BI that does not fuse the representations in a unified one, we first calculate the dispersion of each view and then average them to obtain the overall dispersion). Surprisingly, for almost all of the alignment and dispersion metrics, CrossCBR+BI is better than MultiCBR. This phenomenon reminds us that, in multi-view bundle recommendation, enhancing the cross-view affinity or ego-view dispersion may not be exactly consistent with learning objectives, \aka the user-bundle preference. The cross-view alignment is based on intuition and empirical results, and there is no solid proof about why explicit cross-view alignment can enhance the performance. We discard the cross-view alignment and directly model the cross-view and ego-view preference, demonstrating to be a better solution. 

\begin{table}[t]
\begin{center}
\caption{Performance of MultiCBR under different data augmentations.}
\label{tab:ablation_data_augmentation}
    \resizebox{0.65\textwidth}{!}{
        \begin{tabular}{l | cc | cc}
            \hline
            \multirow{2}{*}{Model} & \multicolumn{2}{c|}{NetEase} &\multicolumn{2}{c}{iFashion} \\
            \cline{2-5}
             & Rec@20 & NDCG@20 & Rec@20 & NDCG@20  \\
            \hline
            \hline
            \textbf{MultiCBR\_OP} & 0.0890 & 0.0483 & 0.1495 & 0.1203 \\
            \textbf{MultiCBR\_ED} & 0.0892 & 0.0484 & 0.1480 & 0.1192 \\
            \textbf{MultiCBR\_MD} & 0.0884 & 0.0479 & 0.1479 & 0.1191 \\
            \textbf{MultiCBR\_Noise} & 0.0909 & 0.0495 & 0.1515 & 0.1223 \\
            \hline
            \textbf{p-value} & 4.10E-02 & 9.48E-04 & 1.10E-08 & 3.16E-12 \\
            \hline
            \hline
            \textbf{CrossCBR} & 0.0844 & 0.0458 & 0.1132 & 0.0872 \\
            \textbf{CrossCBR\_Noise} & 0.0858 & 0.0460 & 0.1139 & 0.0863 \\
            \hline
        \end{tabular}
    }
\end{center}
\end{table}

\subsubsection{Effects of data augmentation} \label{subsubsec:model_study_data_augmentation}
We try various data augmentation methods to generate positive pairs for contrastive learning and present the performances of MultiCBR models using different data augmentation methods in Table~\ref{tab:ablation_data_augmentation}. MultiCBR\_ED represents Edge Dropout, MultiCBR\_MD refers to Message Dropout, and MultiCBR\_Noise corresponds to Noise augmentation. To alleviate the influence of randomness, for each data augmentation, we repeat the model training three times using the fixed best hyper-parameter we found, and take the average performance as the reported value in Table~\ref{tab:ablation_data_augmentation}. \yz{We further adopt ANOVA significance test \cite{girden1992anova} over the performances of MultiCBR using different data augmentations. All of the evaluated p-values are below 0.05, which indicates that data augmentation has a statistically significant influence on MultiCBR's performance.} The results in the table also demonstrate that MultiCBR retains the outstanding performance that is far beyond the baseline performances using all three data augmentations. From the comparison of these three data augmentation methods, noise augmentation gains the best overall performance, suggesting that proper and stronger data augmentations could further enhance the model performance.

To make the comparison fair between MultiCBR and the SOTA method CrossCBR, we also implement the advanced data augmentation method \textit{Noise} to CrossCBR, denoted as CrossCBR\_Noise. The results on NetEase and iFashion datasets are illustrated in Table~\ref{tab:ablation_data_augmentation}. From the results we can observe that CrossCBR\_Noise achieves slightly better or comparable performance with the original CrossCBR model, while still obviously underperforms our proposed MultiCBR. This observation implies that various data augmentation approaches may affect the performance of cross- or multi-view contrastive learning, but just with limited or small scale.

\begin{figure}[t]
    \centering
    \includegraphics[width = 0.9\linewidth]{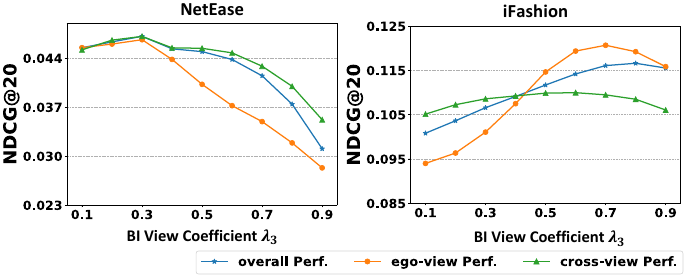}
    \caption{BI view coefficients analysis. The horizontal axis of all the sub-figures is the view coefficient on the BI view, and the vertial axis corresponds to the NDCG@20 score of the model.}
    \label{fig:model_study_2_1}
\end{figure}

\subsubsection{Hyper-parameter Analysis} \label{subsubsec:hyper-parameter_analysis} 
Several key hyper-parameters are crucial for MultiCBR and require carefully tuning to achieve optimal performance. We analyze how our model will perform against the changing of these hyper-parameters, including the BI view coefficient, the contrastive loss weight, and the temperature of the contrastive loss.

\textbf{Effects of view coefficients.} In this paper, we incorporate different view coefficients on each graph learning module to capture the heterogeneity of graphs. Especially, the BI view is newly introduced in MultiCBR and we pay special attention to the corresponding BI view coefficient. To analyze its effect, we showcase the effect of BI view coefficient that is crucial to the performance. Figure~\ref{fig:model_study_2_1} illustrates the performance change in terms of BI-view coefficients. 
\yz{To further investigate the effect of BI view coefficient on ego-view and cross-view preference modeling, we also present the ego-view and cross-view model performance. In Equation~\ref{eq_10}, we just keep the ego-view preference terms and train the model, yielding the ego-view model performance. Analogously, if we just keep the cross-view preference terms and we can obtain the cross-view model performance.
Compared with the case where BI view is discarded ($\lambda_3$ = 0), the performance curves on both datasets demonstrate that BI view is helpful for all of the ego-view, cross-view, and overall performances.} The contribution of BI view differs on different datasets to achieve the best performance, indicating that a good balance between multiple views is crucial for optimal multi-view cooperation. 


\begin{figure}[t]
    \centering
    \includegraphics[width = 0.9\linewidth]{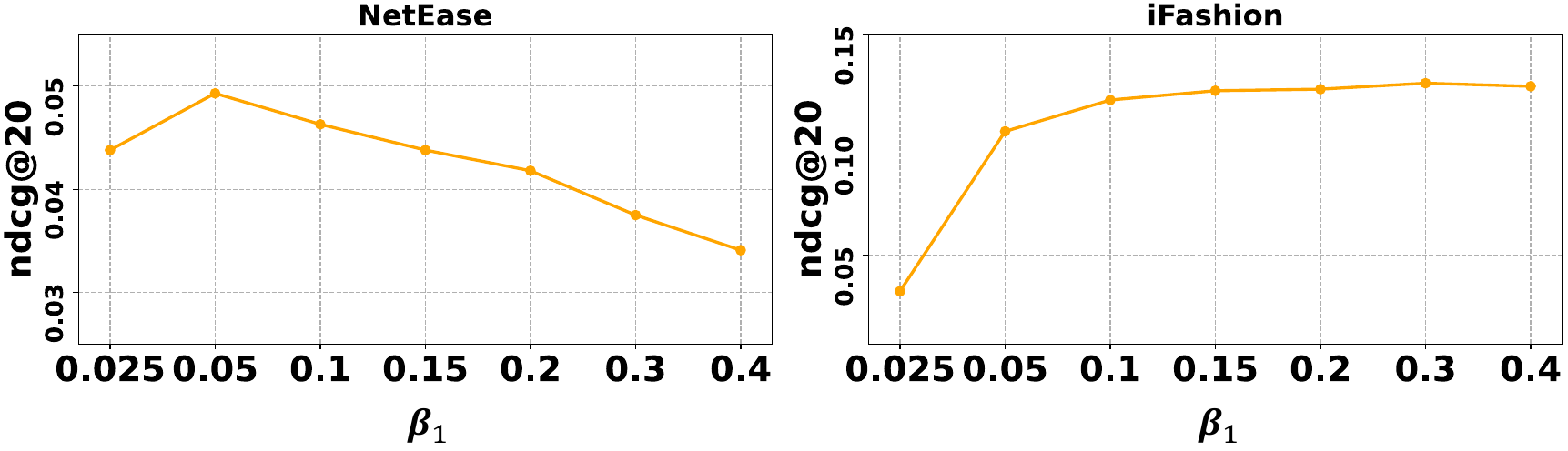}
    \caption{The contrastive loss weight $\beta_1$ analysis. The horizontal axis represents various values of $\beta_1$ and the vertical axis shows the corresponding performance of ndcg@20.}
    \label{fig:model_study_2_2}
\end{figure}

\textbf{Effects of contrastive loss weight $\beta_1$}. Contrastive loss is one of the key components of MultiCBR, and the weight of it may severely affect its performance. We keep all the other hyper-parameters as the optimal values and vary the value of hyper-parameter $\beta_1$ to see the performance change. The results of dataset NetEase and iFashion are shown in Figure~\ref{fig:model_study_2_2}. We have the following observations: 1) on both datasets, the performance varies largely when $\beta_1$ changes, indicating that improper setting of contrastive loss weight may lead to failure of the model; 2) the best setting of $\beta_1$ can be quickly identified via the grid search within a small range, demonstrating that it is easy to tune our model in practice.

\begin{figure}[t]
    \centering
    \includegraphics[width = 0.9\linewidth]{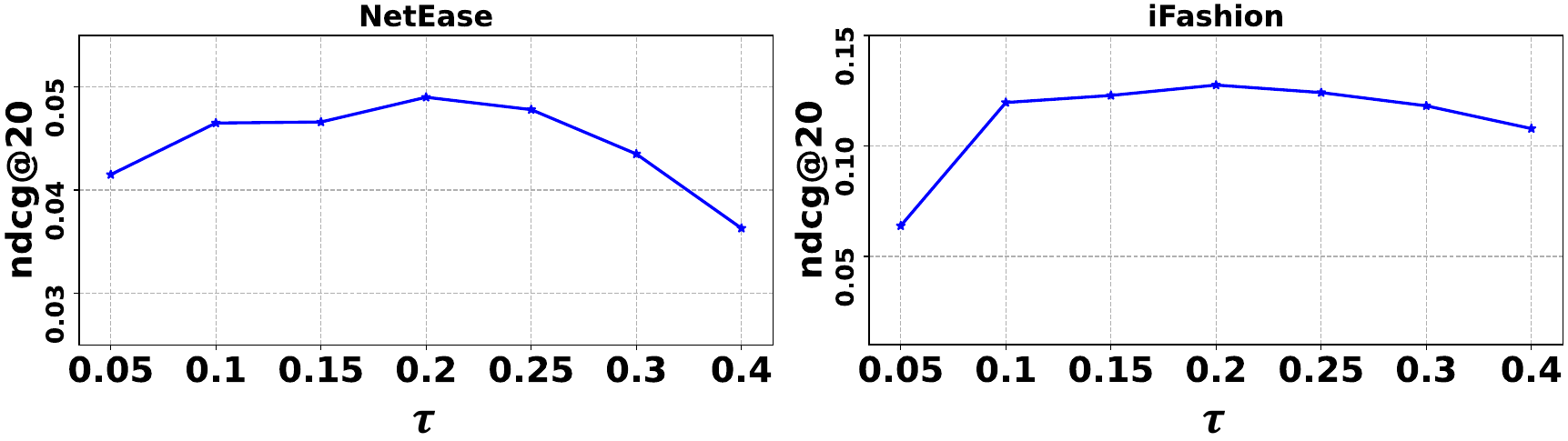}
    \caption{The temperature $\tau$ analysis. The horizontal axis represents various values of $\tau$ and the vertical axis shows the corresponding performance of ndcg@20.}
    \label{fig:model_study_2_3}
\end{figure}

\textbf{Effects of temperature $\tau$}. Another hyper-parameter that most of the contrastive learning-based models are sensitive to is the temperature $\tau$ in Equation~\ref{eq:eq_11}. In order to investigate how temperature $\tau$ impact the model, we fix all the other hyper-parameters as the optimal settings and just change $\tau$. The results illustrated in Figure~\ref{fig:model_study_2_3} demonstrate that MultiCBR is quite sensitive to the hyper-parameter $\tau$ and improper setting of $\tau$ may lead to catastrophic result. Fortunately, analogous to the hyper-parameter $\beta_1$, it is not difficult to obtain the best setting of $\tau$ by grid searching over a restricted candidate set of $\tau$. To be noted, the property of $\tau$ that we observed is similar with the conclusion derived by CrossCBR~\cite{CrossCBR2022}. 

\subsubsection{Computational Efficiency}
To demonstrate the computational efficiency of our model, we record the average running time for one training epoch of our model and two baselines on two devices (\ie Titan V and A5000), as depicted in Table~\ref{tab:training_time}. From the results, we have the following observations. First, MultiCBR is more efficient than CrossCBR+BI since the "early fusion and late contrast" framework only includes two terms of contrastive loss while CrossCBR+BI has six terms. This verifies our hypothesis that the "early fusion and late contrast" design is more efficient than the previous "early contrast and late fusion" strategy. Second, MultiCBR still takes more training time than CrossCBR. As is analyzed in section~\ref{subsec:complexity_analysis}, MultiCBR incorporates one more view than CrossCBR and the graph learning of the additional BI view results in additional costs. Nevertheless, the BI view is necessary to address the BI sparsity issue, thus mainly contributing to the performance improvement.

In addition to analyzing the per-epoch training time, we also record the number of epochs used until the model convergence. As shown in Table~\ref{tab:convergence_speed}, MultiCBR takes the same number of training epochs to converge on the NetEase dataset, while takes fewer number of training epochs on iFashion. Considering both the convergence speed as well as the per-epoch training time, we can conclude that MultiCBR is approaximately as effiecient as the SOTA method CrossCBR.

\begin{table}[t]
\begin{center}
\caption{The efficiency analysis of MultiCBR. Each value is the average running time (seconds) for one training epoch of the model on corresponding dataset.}
\label{tab:training_time}
    \resizebox{0.55\textwidth}{!}{
        \begin{tabular}{c | cc | cc}
            \hline
            \multirow{2}{*}{Models} & \multicolumn{2}{c|}{NetEase} &\multicolumn{2}{c}{iFashion} \\
            \cline{2-5}
             & Titan V & A5000 & Titan V & A5000 \\
            \hline
            CrossCBR    & 5.11 & 3.65 & 30.90 & 26.83 \\
            CrossCBR+BI & 8.92 & 6.96 & 44.86 & 37.53 \\
            MultiCBR    & 8.17 & 6.67 & 41.15 & 36.30 \\
            \hline
        \end{tabular}
    }
\end{center}
\end{table}

\begin{table}[t]
\begin{center}
\caption{The convergence speed comparison of CrossCBR and MultiCBR. Each number is the training epochs used until convergence of corresponding model on certain dataset.}
\label{tab:convergence_speed}
    \resizebox{0.35\textwidth}{!}{
        \begin{tabular}{c | cc}
            \hline
             Models & NetEase & iFashion  \\
            \hline
            CrossCBR & 35 & 20 \\
            MultiCBR & 35 & 25 \\
            \hline
        \end{tabular}
    }
\end{center}
\end{table}
\section{Related Work} \label{sec:related_work}
In this section, we review the related works from three branches: 1) bundle recommendation, 2) graph neural network for recommendation; and 3) contrastive learning for recommendation.

\subsection{Bundle Recommendation}
Bundle recommendation~\cite{RevisitBundle22} has evolved from modeling single user-bundle relation to the modeling of multiple relations among users, bundles and items. FPMC~\cite{FPMC2010} treats each bundle as an atomic object and employs factorization models to model the user-bundle interactions. Later on, people realize that bundle is not an atomic unit, while its included items are crucial for the user-bundle preference modeling. Early works~\cite{cao2017embedding,DAM2019} utilize factorization models or attention mechanism to capture the rich compositional patterns of bundles. In addition, multi-task learning is also utilized to concurrently model both user-item and user-bundle preference, which can enhance the preference modeling from both item and bundle levels. Despite great performance improvement of these works~\cite{cao2017embedding,DAM2019}, the backbone for these models is still based on typical shallow-layered factorization models or attention mechanism, which cannot model complicated higher-order interactions and perform less effectively. 

Recently, GNNs have been employed to capture the complicated relations between users, items and bundles, therefore extensively improve the SOTA performance. Representative methods include BundleNet~\cite{BundleNet2020} and BGCN~\cite{BGCN2020}. BundleNet first builds a tripartite graph among user, bundle, and item to unify all the relations into one graph. It can model the high-order and multi-path interactions over the user-item-bundle tripartite graph via message passing. Different from BundleNet, Chang \etal proposes BGCN~\cite{BGCN2020} that separates the relations among user, bundle, and item into two parts and constructs two levels of graphs, each corresponding to one level of user preference. In particular, it constructs the bundle-level graph based on the user-bundle interactions and constructs the item-level graph based on the user-item interactions and bundle-item affiliations. Thereby, graph neural networks are utilized to learn representations of user and bundle over the two levels of graphs. BGCN not only achieves competitive performance on benchmark datasets but also contributes a novel and compelling approach to modeling bundle recommendation, \ie the two-view formulation to capture both bundle-level and item-level user preferences.

With the emergence of contrastive graph learning, bundle recommendation obtains further enhancements. For example, MIDGN~\cite{MIDGN2022} separates the user-bundle preference into local and global view and then apply contrastive loss in-between the two views. In addition, it also adopts the idea of intent disentanglement to model multiple latent intents for each interaction. MIDGN extensively improves the performance of bundle recommendation on the benchmark datasets, demonstrating the effectiveness of contrastive learning. Concurrently, another contrastive learning-based work, \ie CrossCBR~\cite{CrossCBR2022}, is proposed. It builds two views of bundle and item, following the formulation of BGCN~\cite{BGCN2020}, and then leverages the cross-view contrastive learning to increase the representation affinity of the same user (bundle) while decreases that of different user (bundle). CrossCBR significantly boosts the performance of BGCN and largely improves the SOTA with a simple and efficient graph contrastive learning framework. The success of CrossCBR verifies two main ideas: 1) the bundle and item view formulation is essential in capturing different levels of user preference, and 2) cross-view contrastive learning that explicitly models the cooperative association while enlarging the uniformity of representations is crucial for the performance boosts. In this work, MultiCBR inherits the idea of CrossCBR for view's formulation and further extends the two views formulation to multiple views. MultiCBR also improves the contrastive learning framework to cope with the multi-view scenario. 

In parallel to our formulation of bundle recommendation, there are also some works that are highly related to the topic of bundle recommendation. For example, He \etal extends the problem of bundle recommendation to the scenario of conversational recommendation, thus formulating conversational bundle recommendation~\cite{BudleMCR2022}. Different from recommendation, bundle generation is also a key problem which aims to automatically construct bundle based on existing bundles or user preferences~\cite{BGN2019,BGCN2021,CLHE,TOG2023}. Some other works, such as basket or package recommendation~\cite{basket2020,basket2021,package2016}, release the strict constraint of pre-defined bundle but have a very similar formulation with bundle recommendation, \ie recommending a set of items to users. These works are also popular and meaningful, however they have different settings and formulations with our work. \yz{Basket recommendation recommends a set of items to users, while this set of items is not necessarily a predefined bundle. For bundle recommendation, platforms usually construct some bundles and treat them as a special type of items. Such bundles could be built by sellers (bundles in e-commerce platforms) or users in a crowd-sourcing manner (playlists in the music platform). Thereby, technically, we can have an id for each bundle, while it can hardly have an id for the basket, since each basket does not appear that frequently as bundles.}

\subsection{Graph Neural Network for Recommendation}
Graph neural networks (GNNs) have experienced explosive development in recent years and become the de facto methods to model graph structured data for various tasks, such as graph node classification, link prediction, and graph classification, where representative methods include GCN~\cite{GCN2016}, GAT~\cite{GAT2017}, GraphSage~\cite{GraphSage2017}, and \textit{etc}. As one type of graph structured data, the user-item interactions is naturally suitable for the GNNs and various GNN-based recommendation methods~\cite{NGCF2019,PinSage2018,LightGCN2020} are proposed. He \etal builds a highly-scalable GCN framework, which is based on random walk and can operate on billions of nodes and edges. It is not only a pioneering work that bring GNN to recommender models but also provides practical solutions for the deployment of GNN-based recommender models. Wang \etal formulates the user-item interactions as a bipartite graph and utilizes GCN kernels to learn user and item representations~\cite{NGCF2019}. NGCF achieves great performance on various CF-based recommendation scenarios and applications. Following NGCF, a lighter but more effective version of CF-based GNN model, \ie LightGCN~\cite{LightGCN2020}, is proposed. By removing the feature transformation and non-linear activation function, LightGCN captures the high-order CF signals much better and more efficiently. With the proliferation of contrastive learning, incorporating contrastive learning into the GNN-based recommender models further improve the SOTA performance, which will be introduced in the following section.

Besides typical user-item recommendation, GNNs have also been widely adapted to various sub-problems in recommendation, such as sequential~\cite{chang2021sequential,DGSR2021}, session~\cite{SRGNN2019}, multimedia~\cite{MMGCN2019}, KG-enhanced~\cite{KGAT2019}, and bundle recommendation~\cite{BundleNet2020,BGCN2020}, and \textit{etc}. For example, SRGNN~\cite{SRGNN2019} proposes to construct a graph for each session and employ GCN to capture the transitional patterns of items within each session. To incorporate knowledge graph into recommender models, KGAT~\cite{KGAT2019} integrates knowledge graph into the user-item interaction graph, then employ GNN to learn representations of users and items over the holistic graph. Analogously, GNN is suitable for bundle recommendation, where diverse relations exist among user, item, and bundle. By using GNN to capture the user-item, user-bundle, and bundle-item relations, a list of works have been proposed, such as BundleNet~\cite{BundleNet2020}, BGCN~\cite{BGCN2020}, CrossCBR~\cite{CrossCBR2022}, \etc, as we describe in the above sub-section. In this paper, our model inherits the current merits of GNN in bundle recommendation and further promote this research direction.

\subsection{Contrastive Learning for Recommendation}

Recently, contrastive learning revamps self-supervised learning and achieves enormous progress in areas spanning computer vision~\cite{SimCLR2020,BYOL2020,SimSiam2021}, natural language processing (NLP)~\cite{SimCSE2021}, vision-language tasks~\cite{CLIP2021} and graph learning~\cite{DGI2019,GMI2020,MVGRL2020}, \textit{etc}. For example, SimCLR~\cite{SimCLR2020} presents a simple self-supervised framework to learn visual representations via contrastive learning. It significantly outperforms previous SOTA self-supervised and semi-supervised methods, even matching the performance of a supervised ResNet-50. Inspired by the great success of SimCLR in computer vision, contrastive learning has been quickly adapted to other domains, such as NLP, graph learning, and recommender systems.

Specifically, in the domain of recommender system, contrastive learning has been incorporated into multiple recommendation scenarios, including typical user-item recommendation~\cite{SGL2021,SimGCL2022}, sequential (session) recommendation~\cite{S32020,COTREC21}, cold-start recommendation~\cite{CLC4Rec2021}, cross-domain recommendation~\cite{CCDR2022}, as well as bundle recommendation~\cite{MIDGN2022,CrossCBR2022} \textit{etc}. The main idea of current contrastive learning in recommendation lies in increasing the affinity of positive pairs while decreasing that of negative pairs. The positive pairs are constructed either from stochastic data augmentation~\cite{SGL2021,SimGCL2022,XSimGCL2022,LightGCL2023,S32020} or based on the inherent multiple views of the same node~\cite{CCDR2022,MIDGN2022,CrossCBR2022}. The key of contrastive learning-based recommender models lies in constructing the contrastive pairs. For user-item based recommender models, SGL~\cite{SGL2021} proposes random node dropout, edge dropout, and random walk to generate an augmented view and build a positive pair. SimGCL~\cite{SimGCL2022} and XSimGCL~\cite{XSimGCL2022} finds that a small-scaled random noise is an effective data augmentation method, while LightGCL~\cite{LightGCL2023} introduces SVD as a novel data augmentation method. In sequential recommendation, various sequence augmentation methods, such as dropout, replace, shuffle items in a sequence, have been applied and achieved great performance. In bundle recommendation, MIDGN~\cite{MIDGN2022} and CrossCBR~\cite{CrossCBR2022} first construct two views of representations, then apply contrastive loss over the two views. Even though data augmentation is helpful for performance improvement, the enhancement is marginal according to the report of CrossCBR~\cite{CrossCBR2022}. In this work, we just borrow the data augmentation methods of previous works~\cite{SGL2021,XSimGCL2022} and specifically focus on the order of "fusion and contrast", which is an interesting problem in the multi-view contrastive learning problem.

\section{Conclusion and Future Work} \label{conclusion}
In this paper, we addressed the problem of bundle recommendation with a novel multi-view contrastive learning framework MultiCBR. We constructed a multi-view representation learning framework, which fully exploits all the relational information among users, bundles and items. Especially, in order to alleviate the influence of sparse bundle-item affiliation signals, we introduce graph learning on bundle-item affiliations, resulting in better representation learning and increased performance. Moreover, we adopted an "early fusion and late contrast" strategy, which enhances the user preference modeling through directly modeling both ego-view and cross-view user preference. In addition, compared with previous "early contrast and late fusion" approaches, MultiCBR is more efficient \wrt both computational and optimization costs. We conduct extensive experiments on three benchmark dataset, and the results indicate that MultiCBR outperforms SOTA methods on three public datasets. Especially on the iFashion dataset, our method improves the SOTA over 30\%. We also conduct diverse ablation and model studies to illustrate the working mechanism of MultiCBR, \ie the introduce BI view can counter the BI sparsity issue and the "early fusion and late contrast" is an effective and efficient framework.

Albeit the effectiveness and efficiency of MultiCBR, there remain multiple key problems that deserve further investigation in the future. First, current multi-view fusion still relies on the coarse-level of views with hyper-parameters (\ie view coefficients), however, automatically learning fine-grained user (bundle)-level coefficients for multi-view fusion may achieve better cooperation. Second, our results illustrate that the cosine similarity based cross-view contrastive loss is not always in line with the recommendation objective. Hence, further investigation on the co-effects of contrastive loss and BPR loss is promising to demystify this phenomenon. Third, even though the "early fusion and late contrast" framework has been justified in our setting of three views, it is interesting to consider more views, \yz{such as multimodal features of the items}. Moreover, this novel multiview contrastive learning mechanism can be generalized to other problems, such as cross-domain recommendation and multimedia recommendation.
\yz{Lastly, MultiCBR is a collaborative filtering-based bundle recommendation model that utilizes only relational data. One promising future direction is to introduce multimodal features (\ie images and descriptions of items) and further boost the recommendation quality. Modeling user preference with multimodal features could further alleviate the cold-start problem where relational data is missing for new items in the system.}

\bibliographystyle{ACM-Reference-Format}
\bibliography{references}

\end{document}